\documentclass[a4paper,12pt]{article}

\usepackage{latexsym}

\newcommand{\ud}{\mathrm{d}}

\begin{document}

\title{Stability of a metric $f(R)$ gravity theory implies 
the Newtonian limit}

\author{Leszek M. SOKO\L{}OWSKI \\
Astronomical Observatory and Centre for Astrophysics,\\
Jagiellonian University,\\ 
Orla 171,  Krak\'ow 30-244, Poland \\lech.sokolowski@uj.edu.pl} 

\date{}
\maketitle

\centerline{Short title: Newtonian limit of gravity theories}

\begin{abstract}
We show that the existence of the Newtonian limit cannot work as a selection rule 
for choosing the correct gravity theory from the set of all $L=f(R)$ gravity theories. 
To this end we prove that stability of the ground state solution in arbitrary 
purely metric $f(R)$ gravity implies the existence of the Newtonian limit of the 
theory. And the stability is assumed to be the fundamental criterion of viability of 
any gravity theory. The Newtonian limit is either strict in the mathematical sense 
if the stable ground state of a theory is flat spacetime, or approximate and 
valid on length scales much smaller than the cosmological scale if the ground 
state is de Sitter or anti--de Sitter space. Hence regarding the Newtonian 
limit a metric $f(R)$ gravity does not differ from general relativity (with 
arbitrary $\Lambda$). That stability implies the existence of the 
Newtonian limit is exceptional to Lagrangians depending on $R$ and/or 
the Ricci tensor but not on the Weyl tensor. An independent selection rule is 
necessary.
\end{abstract}

PACS number: 04.50.Kd

\section{Introduction and summary} 
 In recent years the metric nonlinear gravity (NLG) theories have attracted 
vivid attention as a possible explanation for the acceleration of the 
universe without invoking the dark energy concept. These theories differ 
from general relativity only by their Lagrangian $L=f(g_{\mu\nu}, 
R_{\alpha\beta\mu\nu})$ being any smooth scalar function of the 
Riemann--Christoffel tensor for the metric $g_{\mu\nu}$. As the cardinality 
of the set of all analytic functions of one or several variables is higher 
than continuum, the fundamental problem is to choose one or at worst a 
narrow class of Lagrangians out of this set. These theories are tested 
where they are required, i.e.~in cosmology, and exclusively applying the 
Robertson--Walker (RW) spacetime (usually in the special case where the 
spatial sections are flat, $k=0$). However this spacetime is "flexible" in 
that it contains an arbitrary function and thus provides a Friedmannian 
solution for any NLG theory (except some singular cases) while Minkowski, 
de Sitter and anti--de Sitter spacetimes are not universal solutions. This 
implies that there is an infinity of Lagrangians generating solutions 
exhibiting the accelerated evolution of the world. In fact, roughly one 
half of the Lagrangians investigated up to now in the literature fairly 
well fit the astronomical data. Also the solar system tests cannot uniquely 
put stringent bounds on possible Lagrangians due to inherent ambiguity of 
NLG theories. In these theories there is infinity of mathematically 
equivalent dynamical frames out of which only two, Jordan frame and Einstein 
one are employed in practice, and in different frames the initial and 
boundary conditions for the full gravitational field are more or less 
determined by the matter distribution. For example, in Jordan frame the 
initial and boundary conditions are determined by the local matter 
distribution while in Einstein frame they are determined only in a part by 
it \cite{S1}. In this sense the solar system observations are to some 
extent inconclusive.\\

Instead of attempting to deduce the correct NLG theory from the 
astronomical data, which are scarce and theoretically ambiguous, one should 
first verify if the theory (or a class of) under consideration meets the 
general requirements imposed on any classical field theory. A fundamental 
and indisputable criterion is that a theory have a stable maximally 
symmetric ground state. This criterion works effectively and it has been 
shown that many $L=f(R)$ gravity theories ($R$ being the curvature 
scalar for the Riemann tensor) which are attractive on 
cosmological grounds, are actually unstable and thus untenable \cite{S2}. 
Unfortunately still infinite number of Lagrangians is allowed by this 
criterion and further viability conditions are needed to reduce the set of 
tenable gravity theories. Undoubtedly the existence of a properly defined 
Newtonian limit should a priori be such a criterion.\\

The textbook definition of the Newtonian limit of general relativity or an 
alternative gravity theory is that it is a static weak--field limit of 
gravitational interactions corresponding to slow--motion approximation for 
self--gravitating matter systems whose energy--momentum tensor is 
dominated by their energy density. Though intuitively clear, the definition 
is mathematically obscure and incomplete and particularly in the case of NLG theories 
it gives rise to some confusion. To make it clear why we claim in this work that a 
large number of metric gravity theories do have a Newtonian limit we first provide 
a very brief review of what is precisely known about this limit in the case of 
general relativity.\\

In order to express in mathematically rigorous terms the notion of Newtonian limit of 
general relativity (GR) it is necessary to first formulate Newton's gravity theory (NGT) 
in a four--dimensional spacetime formalism. Only within such a wide framework comprising 
both relativistic metric gravity theories (for the time being out of all these theories 
only GR is relevant for our considerations) and NGT proper, where the structure of all 
these theories has the same mathematical and physical interpretation, can one make a 
meaningful transition to Newton's theory and recognize for which relativistic theories 
the 'Newtonian limit' does exist and for which ones does not. The spacetime formulation 
of NGT is as follows.\\
The physical four--dimensional spacetime $M$ is foliated by hypersurfaces $S_t$ of 
simultaneity with respect to the absolute time $t$ and the hypersurfaces are simply 
connected complete Euclidean spaces. The spacetime is endowed with a spatial 
metric\footnote{The Greek indices run from 0 to 3 and Latin ones from 1 to 3.} 
$s^{\alpha\beta}$ of rank 3 which defines the Euclidean metric equal to $\delta_{ik}$ on 
each $S_t$ and a temporal metric $t_{\alpha\beta}$ of rank 1 measuring temporal intervals. 
The curvature tensor of $M$ for the symmetric and metric (for both $s^{\alpha\beta}$ and 
$t_{\alpha\beta}$) connection satisfies the Einstein field equations (with the cosmological 
constant $\Lambda=0$) for the matter source being the energy--momentum tensor for a perfect 
fluid \cite{E}. Actually this theory is more general than NGT and is called Newton--Cartan 
theory. In order to get Newton's theory proper one must impose global and asymptotic conditions 
since Newton's gravity is a theory of isolated material systems and only for isolated 
systems it has been reliably confirmed. The isolatedness is expressed in two conditions:\\ 
i) on each $S_t$ the support of the fluid energy--momentum tensor is compact;\\
ii) the spacetime is asymptotically spatially flat (two expressions quadratic in the 
Riemann tensor vanish at spatial infinity) \cite{Tr, E}.\\
The first requirement means that we are always concerned with isolated systems and the rest 
of the world outside them is empty. The second condition implies that in this special case of 
Newton--Cartan theory there exists a distant parallelism of spatial vectors; in physical 
terms this means that the axes of neighbouring freely falling gyroscopes do not rotate with 
respect to each other. The two conditions explicitly exclude the cosmological constant from the 
field equations. The connection is uniquely determined by a scalar function and the field 
equations reduce to Poisson's equation for the function which is then interpreted as the 
gravitational Newtonian potential. The standard integral formula for the general solution of 
the equation valid in the case of a compact support of the mass density shows that the 
connection falls off as $r^{-2}$ at the infinity.\\
Having this formulation of NGT in hand one places it in a general framework of relativistic 
metric gravity theories. The framework is named Ehlers' frame theory or Cartan--Friedrichs 
formalism \cite{E}. The starting point is GR whose laws are recasted in terms of an arbitrary 
parameter $\mu$ and Einstein's theory is recovered for $\mu=c^{-2}$, $c$ the light 
velocity. The Lorentzian spacetime metric $g_{\alpha\beta}$ gives rise to a temporal metric 
$t_{\alpha\beta}(\mu)$ and spatial one $s^{\alpha\beta}(\mu)$, the connection $\Gamma(\mu)$ 
is symmetric and metric. The field equations are Einstein ones for the connection and the 
matter source is any perfect fluid. The frame theory is meaningful for any $\mu\geq 0$. For 
$\mu>0$ the metrics $t_{\alpha\beta}(\mu)$ and $s^{\alpha\beta}(\mu)$ are of rank 4 while for 
$\mu=0$ they degenerate and the pair $(t_{\alpha\beta}, s^{\alpha\beta})$ forms the so--called 
Galilei metric. All the fields of the frame theory form a set 
$F(\mu)\equiv \{t_{\alpha\beta}(\mu), s^{\alpha\beta}(\mu), T^{\alpha\beta}(\mu),\ldots\}$. 
If the fields belonging to $F(\mu)$ and their first derivatives converge pointwise to those 
forming the family $F(0)$, $\lim_{\mu\to 0} F(\mu)=F(0)$, then $F(0)$ is said to be a 
Cartan--Friedrichs limit of $\mu$--rescaled solutions of GR. $F(0)$ represents laws and 
solutions of Newton--Cartan theory \cite{E}. Again to recover Newton's theory proper the following 
global conditions are imposed in the frame theory:\\
i) the support of $T^{\alpha\beta}(\mu)$ is spatially compact;\\
ii) the spacetime $(M, t_{\alpha\beta}(\mu), s^{\alpha\beta}(\mu), \Gamma(\mu))$ is asymptotically 
spatially flat \cite{Tr, E}.\\
Clearly condition ii) ensures that there are no contributions to the gravitational field from any 
sources outside the isolated system. The tidal gravitational forces due to the system vanish at the 
infinity on each of appropriately chosen spacelike hypersurfaces which foliate the spacetime.\\
It is conjectured that if the fields $F(\mu)$ for $\mu>0$ satisfy the conditions i) and ii) and 
the limit $\lim_{\mu\to 0} F(\mu)=F(0)$ does exist, then $F(0)$ represents the four--dimensional 
formulation of NGT. There is no general proof of the conjecture, only a number of specific 
solutions in GR confirm it \cite{E}. \\
The Ehlers' frame theory shows that the mathematical structure of NGT is a degenerate special 
case of that of GR. This degeneracy may also be expressed in other terms: the spacetime structure of 
GR is determined by a Lorentz bundle over $M$ while that of Newton's theory is given by a 
Galilei bundle \cite{Kue}. These principal fibre bundles are locally determined by their 
structure groups. The Galilei group is a contraction of the Lorentz group showing the degeneracy of 
the structure.\\
It should be emphasized once again that the Newton's theory of gravitation is reliable and the 
Newtonian limit of GR is mathematically well defined only for isolated matter systems, what 
requires asymptotic spatial flatness, otherwise in the limit $\mu\to 0$ one arrives at 
Newton--Cartan theory. For unbounded mass distribution the inertial frames do not exist and the 
structure of NGT is broken. This is the case of cosmology and what is usually named 
'Newtonian cosmology' is a theory which only in some aspects resembles NGT and has no 
evolution equations; evolution of a model follows from some symmetry assumptions and should be 
formulated within the frame theory \cite{RuS}, this means that Newtonian cosmology is not a 
self--contained theory.\\

Strictly speaking for $\Lambda\neq 0$ general relativity has no Newtonian limit. However this is a 
mathematical theorem while in physics one is usually satisfied with a plausible approximation. The 
problem is that of a scale. A physical system determines its own distance scale. In mathematics 
the infinity is unique whereas 'physical infinity' depends on the scale. For example, in 
quantum mechanics the wave function must be normalized to unity over the whole space implying that 
the function must sufficiently quickly fall off at infinity. The scale for the hydrogen atom is 
$10^{-8}$ cm and in practice its infinity is at a distance of few meters and the wave function is 
practically zero there and farther. If the spacetime has a nontrivial topology, is not 
asymptotically flat or there are event horizons and singularities, the rigorous formulation of 
quantum mechanics and quantum field theory encounters the well known difficulties, nevertheless for 
distances small in comparison to the characteristic scale of these features, the quantum phenomena 
are indistinguishable from those in Minkowski spacetime and in this sense standard quantum theory is 
approximately valid. The same holds for gravitational interactions. For the gravitational field of 
the Sun its practical infinity begins not far from the outer edge of the Solar system, i.e.~at 
distance of one parsec. For the Milky Way it is even relatively closer, at the distance of order 
100 Mpc, just outside the outer edge of the Local Supercluster, where the gravitational field is 
dominated by other clusters of galaxies. These scales are small when compared with the characteristic 
scale related to the cosmological constant. The observationally determined upper limit for $\Lambda_0$ 
is $|\Lambda_0|\leq 10^{-52} \textrm{m}^{-2}$ and the characteristic length is at least $10^4$ Mpc 
and is of the order of the Hubble radius $c/H_0$, where $H_0$ is the present value of the Hubble 
constant; the dark energy density corresponding to the upper value of $|\Lambda_0|$ is 
$7\cdot 10^{-30} \textrm{g cm}^{-3}$. Thus both in the Solar system and the Milky Way one can safely 
put $\Lambda=0$. These simple scale comparisons are confirmed by a detailed calculation: the 
cosmological constant is undetectable in the Solar system since the effect most sensitive to it, 
the perihelion shift of Mercury, requires $|\Lambda|\geq 10^{-41} \textrm{m}^{-2}$ \cite{KKL}. If 
the cosmological constant is sufficiently close to zero, small perturbations of de Sitter or 
anti--de Sitter spacetime may be fairly well approximated on macroscopic (i.e.~smaller than 
cosmological) scales by a Newtonian perturbation of Minkowski spacetime.\\

The metric nonlinear gravity theories fit the frame theory but their main problem with 
possessing the Newtonian limit is that they are governed by various fourth order field equations 
while Ehlers' theory requires to this end the Einstein field equations. It might therefore seem 
that only few of them, with very specific field equations, would admit the Newtonian limit. 
However, although for a given NLG theory its field equations are of fourth order in the original 
Jordan frame and in many other frames\footnote{By a 'frame' we always mean in this context a 
set of dynamical variables of the theory. These variables may be subject to arbitrary 
transformations and redefinitions, then the new variables form a new frame.}, there is infinity 
of frames wherein the field equations are of second order. Contrary to a wide belief these 
theories are not inherently dynamically higher order ones. Among the latter frames there is  
one distinguished by the canonical form of its dynamics: it is Einstein frame. We employ this 
frame to show that any $L=f(R)$ gravity theory may be recasted in the form of GR plus a scalar 
field representing a nongeometric spin--zero component of the gravitational field; the scalar 
acts as a 'matter source' for the metric field in Einstein field equations. Then after 
'switching off' this additional gravitational degree of freedom one may directly apply the frame 
theory to get in vacuum the Newtonian limit of the theory under consideration. The procedure 
makes sense if the theory is physically viable, i.e.~if it is stable, what means that its 
ground vacuum state, being either Minkowski, de Sitter or anti--de Sitter spacetime, is dynamically 
stable. Our conclusion is: \textit{if a given metric nonlinear gravity theory has a stable vacuum ground 
state, then it also has a Newtonian limit, either exactly in the sense of the frame theory (if 
the ground state is the flat spacetime) or approximately on a suitable distance scale (if the 
background is curved)}. Stability implies the Newtonian limit. Hence the existence of the 
Newtonian limit cannot work as an independent criterion to establish which NLG theory fits the 
real world. There are still infinity of gravity theories which are in this sense viable and a 
distinct selection rule is necessary to reduce this collection. Such a rule is at present 
missing.\\

In this paper we investigate the Newtonian limit for $L=f(R)$ gravity 
theories, sections 2 to 4. For more general Lagrangians a universal method 
is not available yet and only special cases have been studied; we 
briefly comment on them in section 5. Since the frame theory, being rigorous, is far from the 
physical intuition, for the sake of completeness we discuss in appendix B  
the physical and geometrical obstacles preventing one from defining a Newtonian limit in general 
relativity for $\Lambda\neq 0$, that is as a small perturbation of de Sitter or anti--de Sitter 
spacetime.\\
 We emphasize that our approach and 
results apply to purely metric gravity theories. If one studies $f(R)$ 
gravity in the purely affine or metric--affine framework (Palatini formalism) one may get 
a satisfactory Newtonian limit without invoking a ground state solution 
\cite{BB}.

\section{Stability of a ground state}
We recall that both general relativity (GR) and all metric gravity theories 
differ from Lagrangian classical field theory (CFT) in Minkowski spacetime 
in that the relationship of the notion of energy and of the ground state 
is reverted. In CFT the notion of the ground state is based on the concept 
of energy, being the solution of the Lagrange equations of motion 
corresponding to the lowest energy state; the latter does exist (except 
peculiar cases as Liouville field theory) because the Hamiltonian is 
positive definite. This formal definition agrees with an intuitive picture 
of the ground state in which the field is absent or is covariantly constant 
and its symmetric energy--momentum tensor (derived by taking a formal metric 
variation of the Lagrangian) is either zero or Lorentz invariant.\\
As is well known in GR the Hamiltonian formalism is imperfect and in 
particular cannot be used for defining the ground state. GR is a 
geometrical theory and the ground state is defined in geometrical terms: 
it is the solution of the field equations possessing the maximal 
10--parameter (in dimension four) isometry group, i.e.~admitting 10 
independent Killing vector fields. The solution is unique and depending 
on the value of the cosmological constant $\Lambda$ it is Minkowski ($\cal{M}$), de 
Sitter (dS) or anti--de Sitter (AdS) space\footnote{By 
anti--de Sitter space we always mean the covering AdS space without closed 
timelike curves and with topology $\mathbf{R}^4$.}. The same holds for 
$L=f(R)$ gravity theories, the only difference being that for these 
theories the cosmological constant is not a fundamental one appearing in 
$L$, as we shall see below it is merely the curvature scalar of the 
maximally symmetric solution. And once the primary notion, that of the 
ground state, has been identified in terms of the isometry group, the 
only meaningful notion of energy in GR, that of total energy (being 
effectively a charge) with respect to the ground state, may be introduced 
employing only Einstein's field equations and without any resort to the 
gravitational Hamiltonian. This is Arnowitt--Deser--Misner (ADM) energy 
in the case of $\Lambda=0$ and Abbott--Deser (AD) energy for $\Lambda 
\neq 0$.\\

The fundamental assumption underlying the very notion of the Newtonian 
limit of any relativistic theory of gravity is that the ground state of 
the theory is stable. Otherwise any small time dependent perturbation of 
the Newtonian interaction (which is a specific weak--field solution of 
the relativistic theory) will unboundedly diverge quickly destroying this 
interaction. In short: stability of the background is a necessary condition 
for both viability of the theory and existence of the Newtonian limit.\\

The first step in the search for the Newtonian limit of a gravity theory 
consists in determining the ground state of the theory. For an NLG theory 
with $L=f(R)$ the field equations take on the form (in vacuum) 
\begin{equation}\label{n1}
E_{\mu\nu}(g) \equiv f'(R)R_{\mu\nu} + \frac{1}{6}g_{\mu\nu}[f(R)-
2R f'(R)] - f'''(R)R_{;\mu}R_{;\nu} - f''(R)R_{;\mu\nu} =0,
\end{equation}
here $f'\equiv \frac{df}{dR}$ and we have employed that the trace 
$E_{\mu\nu}g^{\mu\nu}=0$ gives rise to the equation for the scalar $R$,
\begin{equation}\label{n2}
f''(R)\Box R +  f'''(R)R_{;\alpha}R^{;\alpha} + \frac{1}{3}[R f'(R)-
2f(R)]=0,
\end{equation}
where $\Box\equiv g^{\mu\nu}\nabla_{\mu}\nabla_{\nu}$. The ground state 
spacetime should be maximally symmetric, i.e.~Minkowski, de Sitter or 
anti--de Sitter space \cite{Di, S2}. This state 
exists if and only if the field equations admit Einstein spaces, 
$R_{\mu\nu}=\frac{1}{4}\lambda g_{\mu\nu}$ with $R=\lambda =\textrm{const}$, 
as a special class of solutions. From (1) or (2) one finds that the 
curvature scalar $\lambda$ satisfies the algebraic equation \cite{BO, 
Di, S2}
\begin{equation}\label{n3}
\lambda f'(\lambda) - 2 f(\lambda) =0.
\end{equation}
In general eq. (3) has multiple solutions giving rise to multiple vacua 
\cite{HOW}. We exclude from considerations the degenerate Lagrangians 
for which any value of $\lambda$ is a solution of (3) (continuous spectrum, 
$L=R^2$), the only solutions are infinite ($L=1/R$) or the equation has 
no solutions at all \cite{S2}. Thus eq. (3) has at least one and at most 
countable number of finite solutions. Each ground state defines a separate 
dynamical sector of the theory, i.e.~a given Lagrangian corresponds to a 
number of distinct dynamical sectors, each sector being actually a 
distinct gravity theory (for examples cf. \cite{S2}). Classically there 
are no transitions between different sectors for the same Lagrangian; may 
be distinct vacua are related via quantum tunnelling processes. \\

Any root of this equation may be interpreted in a restricted sense as a 
cosmological constant of the theory, $\Lambda \equiv \lambda/4$. In fact, 
a small perturbation of the corresponding ground state has $R$ close to 
$\lambda$ as is the case of general relativity where the ground state 
curvature is $R=4\Lambda$. It is therefore necessary to make a comment on the notion 
of the cosmological constant. In general relativity $\Lambda$ is both the 
constant appearing in the Einstein--Hilbert Lagrangian, $\Lambda=-\frac{1}{2}
L(0)$, hence it explicitly appears in the fields equations and in all solutions to 
them, and the curvature of the unique maximally symmetric ground state, $\Lambda=
\lambda/4$. Yet in metric NLG theories this notion has a very limited sense. If 
$f(0)$ is infinite, as is in most Lagrangians employed in cosmological applications, 
the definition is meaningless. If $f(0)\neq 0$ is finite one may define $\Lambda$ 
as $-\frac{1}{2}f(0)$, however this quantity does not appear in the field 
equations and influences the solutions only in an implicit way via dimensional 
cosntants which are unavoidable to ensure the correct dimensionality of the 
Lagrangian; e.g.~for $f(R)=\frac{1}{a}e^{aR}$ with $a>0$ the unique ground state solution 
is dS space with $\lambda=2/a$. Clearly for $f(0)=0$ this definition gives $\Lambda=0$ 
while in general besides Minkowski space ground state ($\lambda=0$) there are other 
ground states with $\lambda\neq 0$. For example, for $f(R)=R+aR^2+\alpha^{-2}R^3$, 
$\alpha>0$, there are three ground state solutions with $\lambda_1=0$, $\lambda_2=
\alpha$ and $\lambda_3=-\alpha$, defining three separate dynamical sectors of the 
theory. Alternatively, $\Lambda$ may be defined as $\lambda/4$ for each ground state, 
then it has different values in different sectors of the theory. In what follows we shall always 
use the notion of $\Lambda$ only in the sense of the curvature of the (stable) 
maximally symmetric ground state. It is relevant in that solutions to the field 
equations may asymptotically tend to the ground state with $R=4\Lambda$.\\

A given solution to (3) is a genuine ground state of a gravity theory if 
it is stable in this theory against purely gravitational excitations (no 
matter). In the presence of some kind of matter the candidate ground 
state may be stable or not. If some species of matter causes instability, 
this is either an indication that this species is merely unphysical or that 
producing it would be unreasonable and dangerous. Quantum massless fields 
make both Minkowski and de Sitter space unstable \cite {HTH} in general 
relativity and this outcome is not regarded as an argument against 
validity of Einstein's theory; general relativity may be challenged only 
on completely different grounds. What is relevant is the stability of pure 
gravity theory. \\

In general relativity the flat spacetime $\cal{M}$ is globally dynamically 
stable \cite {CK} and de Sitter  space ($\lambda = 4\Lambda>0$) is 
globally nonlinearly stable too \cite{F1}; the case of anti--de Sitter 
space is distinct and we comment on it in Appendix B.\\
In an NLG theory with arbitrary $L=f(R)$ one may investigate the stability 
employing the remarkable fact that general relativity plus a minimally 
coupled scalar field is a universal Hamiltonian image of any such gravity 
theory under a suitable Legendre map \cite{MFF, MS1, S1, S2}. One may 
therefore apply the methods developed in general relativity. The 
classical method is based on positivity of total ADM energy for both 
gravitational field and a matter source. The energy is positive provided the 
energy--momentum tensor for the matter source satisfies the dominant 
energy condition (DEC) and the latter holds if the interaction potential is 
nonnegative and attains minimum at the ground state under consideration. Thus 
investigation of extrema of the potential for the scalar field becomes an 
effective method for studying the stability in NLG theories. \\
It should be noted that from the rigorous mathematical approach viewpoint the 
classical method of proving stability based on the positivity of energy, is of 
rather little reliability \cite{An1}. In proving the dynamical stability (of 
evolution, meaning that there are no unboundedly growing modes) only the exact 
field equations are relevant. However in the few cases in the rigorous approach 
where matter sources are present, DEC does hold. It is therefore reasonable to 
conjecture that $\cal{M}$, dS and AdS are globally nonlinearly stable only if 
any self--gravitating matter (in the present case the scalar component of 
gravity) does satisfy the condition.\\

In Jordan frame (JF) the $L=f(R)$ gravity is described by the field 
$g_{\mu\nu}$ which is a kind of a unifying field mixing the pure spacetime metric 
(still equal to $g_{\mu\nu}$) and a spin--0 component of gravity since the 
unifying field carries 3 degrees of freedom. The field is decomposed into 
the components carrying definite masses and spins in Einstein frame (EF); 
in the latter frame it is a doublet $\mathrm{EF}=\{\tilde{g}_{\mu\nu}, \phi\}$. 
The transformation from JF to EF is a Legendre map being in this case a 
conformal rescaling of the original metric \cite{MFF, MS1, S2}. The scalar 
component of gravity is defined as $p\equiv \frac{\ud f}{\ud R}$, then the 
definition is inverted to give $R$ as a function of the canonical 
momentum $p$, $R(g)=r(p)$, i.e., 
\begin{displaymath}
f'(R)\big|_{R=r(p)} \equiv p.
\end{displaymath}
For convenience the scalar is redefined as 
\begin{displaymath}
p\equiv \exp\left(\sqrt{\frac{2}{3}}\kappa \phi\right)
\end{displaymath}
where $\kappa^2 =8\pi G/c^4$ and the Einstein frame metric is 
$\tilde{g}_{\mu\nu}\equiv p g_{\mu\nu}$. The fourth order field 
equations (1)--(2) in JF are equivalent in EF to 
$\tilde{G}_{\mu\nu}(\tilde{g}) = \kappa^2 T_{\mu\nu}(\phi, \tilde{g})$ 
for a minimally coupled scalar field with a self--interaction 
potential 
\begin{equation}\label{n4}
V(p(\phi)) = \frac{1}{2\kappa^2}\left[\frac{r(p)}{p} -
\frac{f(r(p))}{p^2}\right]
\end{equation}
and the equation of motion 
\begin{equation}\label{n5}
\stackrel{\sim}{\Box}\!\!\phi = \frac{\ud V}{\ud \phi} =
\sqrt{\frac{2}{3}}\kappa p \frac{\ud V}{\ud p}.
\end{equation}
The Legendre transformation to EF should exist at least in a neighbourhood 
of a ground state solution with $R=\lambda$; it occurs iff $f'(\lambda) 
\neq 0$ and $f''(\lambda)\neq 0$. Without loss of generality we assume 
$f'(\lambda)>0$ to preserve the metric signature. The two conditions 
additionally restrict the class of allowable Lagrangians \cite{S2}.\\

The following proposition holds \cite{S2}: \emph{If} 
\begin{displaymath}
\frac{1}{f''(\lambda)} - \frac{\lambda}{f'(\lambda)}>0
\end{displaymath}
\emph{the maximally symmetric solution of the theory for $R=\lambda$ 
is stable against gravitational (i.e.~metric and the scalar field)  
perturbations\footnote{An almost equivalent stability criterion based on 
linear perturbation theory of RW spacetimes in a fourth--order theory in 
Jordan frame has been given in \cite{Fa}. The only difference is that 
the strong inequality  in this formula is replaced by a weak one there.}}. \\
Here some comments are in order.\\
1. The proposition is explicitly formulated in EF and states the stability 
of the ground state in the framework of general relativity. The inverse 
Legendre map is simply 
$g_{\mu\nu}= \frac{1}{f'(\lambda)}\tilde{g}_{\mu\nu}$, hence Minkowski, 
dS and AdS spaces in EF are mapped onto $\cal{M}$, dS and AdS spaces in 
JF respectively, preserving the sign of the curvature scalar. Since it is 
assumed that the Legendre map is regular in a neighbourhood of the ground 
state in EF, the corresponding ground state solution in JF is stable as 
well. We stress this mathematically obvious fact since there were some 
suggestions in the literature that the stability might be spoiled under 
transformation between different frames. \\
2. The proposition is of mathematical nature and its validity is independent 
of the issue of which frame is physical. While considering $f(R)$ gravity 
one is usually interested in physics in Jordan frame, but in this frame the 
stability problem is hard. Yet Einstein frame, which is mathematically 
(though not physically) equivalent to JF, allows to solve the problem in a 
neat and general way. \\
3. Long ago a paper by Pechlaner and Sexl \cite{PS} made impression that 
fourth order equations of motion generate instabilities which are revealed 
whenever a small amount of matter is present \cite{Di}. Actually the higher 
order terms in an equation only signal the presence of additional field 
degrees of freedom; in the present case this is the scalar component of 
gravity. It is the precise form of the full Lagrangian (in JF) rather 
than the mere presence of higher derivatives in the field equations that determines whether 
the ground state is stable or not.\\
4. Here stability means the dynamical stability (linear or exact) of the 
ground state solution of a theory. Yet in the view of the obvious 
cosmological applications, most research in $f(R)$ gravity have up to now 
been focussed on stability of RW spacetimes or other cosmological models 
using either the phase space method or a minisuperspace approach (in both 
the cases the perturbations are spatially homogeneous) \cite{TT} or a 
linear theory of inhomogeneous perturbations \cite{Fa, LB}. However a 
cosmological solution only exceptionally coincides with the ground state 
one\footnote{Recall that a RW spacetime reduces to the flat one 
only for the flat or open spatial sections and and then only provided 
that the cosmic scale factor is constant or a linear function of the 
cosmic time respectively; these trivial cases are not studied in these 
works. Anti--de Sitter space cannot at all be expressed in terms of the 
RW metric, therefore these three approaches do not apply to it.} and we 
stress that it is the stability of the latter that is relevant for the 
physical viability of the theory and for possible existence of a Newtonian 
limit. Investigating solely the cosmological solutions is in a sense 
misleading: as a number of specific examples show, a stable cosmological 
solution may have interesting features \cite{LB} while the ground state 
is unstable making the underlying theory unphysical \cite{S2}. The 
approach based on positivity of total energy (in both the frames!) or more 
precisely, on DEC for the scalar field, is 
universal and in this sense is superior to the other methods.

\section{Gravitational vacuum in Einstein frame and the Newtonian limit}
Assume that a given Lagrangian $L=f(R)$ admits $n$ different solutions of the ground state 
equation (3) $\lambda_i$, $i=1,\ldots,n$, and each of the corresponding maximally 
symmetric spacetimes with $R=\lambda_i$ is stable; this means that the Lagrangian 
describes $n$ physically distinct gravitational sectors of the theory. Consider the 
field equations in Einstein frame. These are $\tilde{G}_{\mu\nu}(\tilde{g}) = 
\kappa^2 T_{\mu\nu}(\phi, \tilde{g})$ and the nonlinear wave equation (5) for $\phi$. 
Clearly an arbitrary solution to these equations cannot have the Newtonian limit. 
In the absence of ordinary matter the 
scalar gravity acts as a specific matter source for the metric field 
$\tilde{g}_{\mu\nu}$ and any solution contains contributions from the 
scalar which are also present in the weak--field limit and perturb the 
Newtonian interaction. 
Therefore in the search for a Newtonian limit one needs to study
"scalar gravity vacuum" solutions where the spin--0 component of gravity is 'switched off', 
i.e. $\phi=\textrm{const}$. This may occur only for a stationary point of the potential 
$V(p(\phi))$ in eq. (5). From the form (4) of the potential one easily finds that 
$\ud V/\ud p=0$ implies $\frac{2}{p}f(r(p))-r(p)=0$. Recalling that $p=f'(r)$ one gets 
that all stationary points are determined by 
\begin{displaymath}
2f(r(p))-r(p)f'(r)=0
\end{displaymath} 
and this equation viewed as an equation for $r(p)$ coincides with eq. (3). Hence 
$\phi=\textrm{const}$ only for $r(p)\equiv r(p_i)=\lambda_i$ and conversely 
$p_i=p(r_i)=p(\lambda_i)=f'(\lambda_i)$. The scalar field is in its ground state either 
in the ground state of the entire gravitational doublet (the spacetime is $\cal{M}$, dS 
or AdS in both JF and EF, depending on the sign of $\lambda_i$) or in a spacetime which 
in Jordan frame has the same curvature scalar $R=\lambda_i$ as the ground state of the 
given sector. Since by assumption $f'(\lambda_i)>0$, in general the scalar gravity does not 
vanish in its ground state, 
\begin{displaymath}
\phi_i=\sqrt{\frac{3}{2}}\,\frac{1}{\kappa}\ln f'(\lambda_i).
\end{displaymath} 
The energy--momentum tensor for $\phi$ reduces to its potential part, 
$T_{\mu\nu}(\phi_i, \tilde{g})=-\tilde{g}_{\mu\nu}V(p_i)$. Also the potential does not 
vanish and from (4) one gets
\begin{displaymath}
V(p_i)=\frac{1}{2\kappa^2}\left[\frac{\lambda_i}{f'(\lambda_i)}-\frac{f(\lambda_i)}
{(f'(\lambda_i))^2}\right]
\end{displaymath} 
and applying eq. (3) for $\lambda=\lambda_i$ and following from it relation 
$f(\lambda_i)/f'(\lambda_i)=\lambda_i/2$ one finally arrives at 
\begin{equation}\label{n6}
V(p_i)=\frac{\lambda_i}{4\kappa^2 f'(\lambda_i)}.
\end{equation}
The Einstein field equations for $\tilde{g}_{\mu\nu}$ may be written in the case 
$\phi=\phi_i$ as 
\begin{equation}\label{n7}
 \tilde{G}_{\mu\nu}(\tilde{g})+\Lambda_i\,\tilde{g}_{\mu\nu}=0
\end{equation}
where 
\begin{equation}\label{n8}
\Lambda_i\equiv \frac{\lambda_i}{4 f'(\lambda_i)}
\end{equation}
is interpreted as a cosmological constant in the given sector of the theory for this 
class of solutions. One sees here another difference betweeen the two frames: while in 
JF the cosmological constant refers only to the curvature of the ground state, in EF 
it also appears in the field equations. (This constant may be singled out in the field 
equations in general, i.e.~when the scalar field is present.)\\
Clearly the gravity theory (7) has the correct Newtonian limit (exact or approximate). 
Coming back to Jordan frame one simply rescales the metric by the constant factor, 
$g_{\mu\nu}=(f'(\lambda_i))^{-1}\,\tilde{g}_{\mu\nu}$, and the gravitational interaction 
takes on the same Newtonian form in this frame too.\\

\section{Minkowski space as the ground state}
Finally we make some remarks about those $L=f(R)$ gravity theories which 
have the flat spacetime as the stable ground state 
solution\footnote{More precisely, we now consider the $\lambda =0$ sector 
of a given theory.}. 
For $\lambda =0$ eq. (3) implies $f(0)=0$ excluding a constant from the 
Lagrangian. Assuming analyticity\footnote{Actually it is necessary to 
assume that $f(R)$ is of $C^3$ class at $R=0$.} and normalizing $f'(0)$ 
to 1 one has\footnote{We use all the conventions of the book \cite{HE}.}
\begin{equation}\label{n9}
L=f(R) = R + aR^2 + \sum_{n=3}^{\infty} c_n R^n 
\end{equation}
with $a\neq 0$. The existence of the first two terms in the expansion 
is essential for the equivalence of Jordan and Einstein frames. The 
conformal factor, i.e.~the scalar component of gravity, is 
\begin{displaymath}
p=1 + 2aR + O(R^2)
\end{displaymath}
and hence is positive in a neighbourhood of the ground 
state\footnote{We notice in passing that, contrary to what is frequently 
met in the current literature on $f(R)$ cosmology, the condition 
$R\approx 0$ does not necessarily imply that the gravitational field is 
weak; as a matter of fact this occurs mainly in the cosmological 
setting of general relativity. In general one may have $R=0$ for 
arbitrarily strong gravity.}. The stability condition reduces now to 
$a>0$ \cite{MS1}. 
The "scalar gravity vacuum" simplifies to  $\phi =0$ or $p=1$ and implies 
$T_{\mu\nu}=0$. Then the 
theory becomes identical to vacuum general relativity and applying 
Ehlers' frame theory to appropriate solutions to $\tilde{G}_{\mu\nu}
=0$ (e.g.~Schwarzschild one) one gets the desired correct Newtonian 
limit of the theory. By taking the inverse conformal mapping one finds 
$g_{\mu\nu}=\tilde{g}_{\mu\nu}$ in JF and thus $G_{\mu\nu}(g)=0$ and 
the same solutions give rise in the weak--field limit to the Newtonian 
interaction. In summary, any theory of the form (9) and $a>0$ has the 
flat spacetime as a stable ground state solution and exactly the Newtonian 
limit interaction described by a potential $U$ satisfying $\Delta U =
0$.\\

It is difficult to derive this theorem working solely in Jordan frame. 
Firstly, there is the problem of proving stability of the ground state. 
Secondly, when one finds out the stability criterion $f''(0)>0$ 
employing perturbation theory as in \cite{Fa}, there remains to decouple 
the massive scalar field contribution from the pure massless gravitation, 
both contained and mixed in the unifying field $g_{\mu\nu}$. (Recall 
that the Newtonian interaction is a far distance force and the very 
presence of the scalar gravity will distort it.) It is impossible to 
decouple the scalar directly from the field equations (1)--(2), one 
can only identify its contribution to specific solutions. \\
Exact solutions (in any frame) for the analytic Lagrangians are not 
known\footnote{Few static spherically symmetric (SSS) solutions 
different from Schwarzschild's one are known in non--analytic cases: 
for $L$ containing $\sqrt{R}$ term \cite{MV} and for $L=R^s$, $s$ real 
\cite{CB, MV, CST1}. If one assumes $R=0$ the unique SSS solution is 
Schwarzschild's metric \cite{CST2}. It is worth noting that a SSS solution 
different from Schwarzschild's one can be found for $L$ being a non-analytic 
function of the Weyl tensor \cite{DST}.}. Of course the Schwarzschild 
metric is always a solution and in the sense of Ehlers' frame theory it 
guarantees the existence of the Newtonian limit, but the physical 
interpretation of the metric (absence of the scalar gravity) cannot be 
recognized on the level of the fourth order equations, i.e.~in Jordan 
frame. In the linear approximation a SSS solution for $L = R + 
aR^2$ was found long ago by Pechlaner and Sexl \cite{PS} and Stelle 
\cite {St}. The general solution to the field equations (1) and (2) 
for the metric components $g_{00}$ and 
$g_{11}$ depends on 3 parameters (before imposition of a boundary 
condition) and for $a>0$ is of the form (up to signs) 
\begin{displaymath}
1 +\frac{c_1}{r} + \frac{c_2}{r}e^{-mr} + \frac{c_3}{r}e^{mr}, 
\end{displaymath}
the sum of the Newtonian and Yukawa potentials. $m=(6a)^{-1/2}$ is 
easily identified in EF  as the mass of the scalar gravity. The 
analogous result was recently found in the linear approximation for 
the general Lagrangian (9) \cite{CST3}. Thus if the scalar degree of 
freedom is switched off ($c_2 =c_3 =0$) the Newtonian interaction is 
a weak--field limit of any SSS solution for each analytic Lagrangian. 
For $a<0$ the approximate solution is complex in the Yukawa terms (the 
mass is imaginary) and 
its physical interpretation given in \cite{CST3} is rather obscure or, 
if only its real part is taken into account, it quickly oscillates \cite{PS}. 
Actually this behaviour just signals the instability (the corresponding 
time--dependent modes are divergent) which is immediately recognized 
in Einstein frame.\\

At first sight the theorem that the stability implies the Newtonian limit 
is perhaps a little surprising. One might a priori imagine gravity theories 
whose ground state is stable under, say, radiative mode transmission while 
they do not admit the Newtonian interaction. This may occur for some 
specific theories while for $L=f(R)$ Lagrangians it is impossible. As 
already mentioned in Section 2, this class of gravity theories is 
distinguished in the entire space of possible relativistic theories of 
gravitation by the fact that they can be Legendre transformed into GR plus 
the scalar field. As long as the ordinary matter is not included, these 
theories merely represent general relativity in disguise.\\
In summary, those $L=f(R)$ theories where flat spacetime is unstable, are 
rejected as unphysical and those for which this spacetime is stable contain 
as a subclass of solutions all the solutions of vacuum general relativity 
(both in Jordan and Einstein frames). This subclass (and only this one) 
contains solutions which subject to the specific mathematical procedure 
give rise to rigorous Newton's gravity theory.
 
\section{More general Lagrangians}
The Legendre transformation from Jordan to Einstein frame works for 
all Lagrangians $L=f(g_{\mu\nu}, R, R_{\alpha\beta})$ (no dependence 
on the Weyl tensor) and whose Hessian with respect to $R_{\alpha\beta}$ 
does not vanish \cite{MFF}. In EF the unifying field $g_{\mu\nu}$ is 
decomposed into a metric $\tilde{g}_{\mu\nu}$ (which now is not 
conformally related to $g_{\mu\nu}$), a massive scalar field $\chi$ 
and a massive spin--two field $\phi_{\mu\nu}$ actually being a 
"ghost" (what is not so disastrous as it might seem, cf. 
\cite{HH}). The indefiniteness of the energy--momentum tensor for 
$\phi_{\mu\nu}$ causes that the general--relativistic method of 
studying stability of the ground state (based on DEC) does not work. 
One can only study various Lagrangians case by case. There are 
arguments that the most physically interesting case corresponds to the 
simplest regular Lagrangian in this class,
\begin{equation}\label{n10}
L= R + \frac{1}{3m^2}(R^2 - 3R_{\mu\nu} R^{\mu\nu}).
\end{equation}
The coefficients are so chosen that the scalar gravity vanishes and the 
gravitational field is a doublet consisting of two spin--2 fields 
carrying together seven degrees of freedom \cite{MS2}; $m$ is the mass 
of the non--metric component of gravity ($\phi_{\mu\nu}$). The ground 
state in both the frames is Minkowski spacetime (in EF it is 
supplemented by $\phi_{\mu\nu}=0$) and is linearly stable \cite{MS2}, 
hence the ghost--like nature of the massive gravity does not result in 
instability. Metrics satisfying $R_{\mu\nu}=0$ are always solutions to 
(10) and this is sufficient to conclude that the theory has the correct 
Newtonian limit. (No exact SSS solutions different from Schwarzschild 
metric are known.)\\

Finally we comment on theories explicitly depending on the Weyl tensor. 
In this case Einstein frame does not exist \cite{MS3} and all frames obtained via 
various Legendre transformations from the original Jordan frame give rise 
to fourth--order equations of motion what makes investigations of these 
theories rather hard. In the special case of 
\begin{equation}\label{n11}
L= R + \sqrt{3}a|C_{\alpha\beta\mu\nu}C^{\alpha\beta\mu\nu}|^{1/2},
\end{equation}
$a<1/4$ or $a>1$, the Lagrangian is a homogeneous function of order 1 
of the Riemann tensor, as in general relativity. For this theory all 
exact SSS solutions have been found \cite{DST} and they do not include 
Schwarzschild metric. A preliminary calculation shows that the 
maximally symmetric spaces, i.e.~flat, dS and AdS spaces are not 
solutions too (though it needs a deeper investigation) and it is unclear 
whether a ground state may at all be defined. If it cannot it would be 
a clear indication that the Lagrangian (11) is unphysical.
 
\section{Conclusions}
As regards the existence of the Newtonian limit the $L=f(R)$ gravity theories 
do not differ from general relativity. If their ground state solution, being 
Minkowski, de Sitter or anti--de Sitter spacetime, is 
dynamically stable (as is the case of general relativity with arbitrary cosmological 
constant), then the Newtonian limit does exist. The limit is either rigorously defined 
in the case $\Lambda=0$, i.e.~for the flat ground state solution, or approximate 
otherwise, the approximation is valid for length scales small compared to the 
cosmological scale being of the order of $|\Lambda|^{-1/2}$. 
Stability of the ground state is a necessary 
and sufficient condition for the Newtonian force to exist, at least on macroscopic 
scales. This unexpected theorem is due to the fact that in vacuum this class of 
gravity theories is dynamically equivalent to general relativity 
(plus a scalar field).\\

For more general Lagrangians depending on Ricci and Weyl tensors the 
situation is more complex. Stability of a ground state spacetime remains the 
necessary condition for the Newtonian limit to exist also in this case but now 
there is no universal, effective and simple method for checking the 
stability and one must resort to perturbation theory. Next, one must 
show that the fourth--order field equations (in the case of Weyl tensor) of the 
theory admit exact solutions having the appropriate Newtonian limit. 

\section*{Acknowledgments}
I am grateful to Michael Anderson, Piotr Bizo\'n, Piotr Chru\'sciel, Zdzis\l{}aw Golda, 
Andrzej Staruszkiewicz and Andrzej Trautman for extensive 
discussions, helpful comments and explanations.

\section*{Appendix A. Problems with solutions in general relativity possessing the exact 
Newtonian limit}
In the two appendices we discuss some problems with obtaining the Newtonian limit in general 
relativity for both $\Lambda=0$ and $\Lambda\neq 0$. Most of the material is by no means new, 
but for the sake of conceptual completeness and the reader's convenience we present it here. In 
this appendix we deal with the case $\Lambda=0$, then there exist exact relativistic solutions 
giving rise to the rigorous Newtonian limit.\\
In Ehlers' frame theory the Newtonian limit 
is defined as a limit for particular classes of solutions (families 
of spacetimes) of GR depending on some free parameters. As the specific examples 
in the third reference in \cite{E} show, the method of the frame theory works 
properly under two assumptions. Firstly, a foliation of the spacetimes by 
spacelike hypersurfaces must be chosen in such a way that after performing 
their linearization the linearized solutions become perturbations of the 
ground state (flat Minkowski spacetime) and the time coordinate labelling 
the foliation becomes the time coordinate in the global inertial reference 
frame being the (almost) proper frame for the matter source. Geometrically 
the latter feature means that the foliating hypersurfaces flatten upon the 
linearization. This time coordinate of the proper inertial frame in 
Minkowski spacetime is then identified with the absolute time in Galilei 
spacetime. Secondly, as the case of the FLRW cosmological spacetime shows 
\cite{E}, a coordinate system compatible with the foliation should be 
carefully chosen, otherwise the derived Newtonian limit of the relativistic 
spacetimes does not resemble at all the ordinary Newton's gravity. In fact, 
in the standard (Lagrangian) Friedmann coordinates in the Newtonian limit 
the gravitational field strength is space independent (and time dependent) 
and the "Newtonian scalar potential" does not exist. (Physically this 
is reasonable showing that the infinite homogeneous distribution of matter, 
static or evolving, is beyond the scope of Newton's gravity theory.)\\
The first condition is crucial. For example, for Schwarzschild spacetime 
expressed in the standard coordinates $(t,r,\theta, \phi)$ where $t$ is 
the parameter on the trajectories of the timelike Killing vector $\partial
/\partial t$, one gets in the Newtonian limit the gravitational 
acceleration $1/r^2$, while after a coordinate transformation 
\begin{displaymath}
t =T \cosh R, \qquad r=T\sinh R,
\end{displaymath} 
the 3--spaces become hyperboloids with the variable curvature scalar. If 
one takes the Newtonian limit in these coordinates following the Ehlers' 
prescription, the temporal and spatial metrics describing Galilei spacetime 
do not acquire their proper forms and the two conditions (given in \cite{E}) 
for convergence of a relativistic class of spacetimes to Newtonian gravity 
are not satisfied; actually the Newtonian limit for this foliation does not 
exist. This result is easily understood: Schwarzschild spacetime in $(T, R)$ 
coordinates reduces in the limit of vanishing mass to flat spacetime 
foliated with the hyperboloids of constant negative curvature (Lobatchevski 
spaces) and this is a fully relativistic description of the spacetime. Yet 
Schwarzschild spacetime in Painlev\'e--Gullstrand coordinates is foliated 
by flat spaces and the metric is time independent and in the limit of 
vanishing mass this metric reduces to that in an inertial frame in flat 
spacetime (i.e.~the foliating spaces become hyperplanes); for this metric 
the Ehlers' method works well giving rise to the correct Newtonian limit. 
It is then essential to properly identify the appropriate foliation of the 
spacetime. \\

Furthermore there are mathematical subtleties causing that 
in general a linearized form of an exact solution written in arbitrary 
parameterization need not be a solution to the linearized field equations. 
In fact, in general relativity (for $\Lambda=0$) one linearizes the Einstein 
field equations around the flat spacetime writing in Cartesian coordinates 
$g_{\mu\nu}= \eta_{\mu\nu} + h_{\mu\nu}$ and assuming that both 
$|h_{\mu\nu}|\ll 1$ and the derivatives $|h_{\mu\nu,\alpha}|\sim 
|h_{\mu\nu,\alpha\beta}|\sim |h_{\mu\nu}|$. Then in the harmonic gauge for 
$\bar{h}_{\mu\nu}=h_{\mu\nu} -\frac{1}{2}\eta_{\mu\nu}h^{\alpha}{}_{\alpha}$ 
one gets the equations 
\renewcommand{\theequation}{\Alph{section}.\arabic{equation}} 
\setcounter{section}{1}
\setcounter{equation}{0}
\begin{equation}
\Box \bar{h}_{\mu\nu} = -16 \pi G T_{\mu\nu}
\end{equation}
with appropriately linearized $T_{\mu\nu}$. This means that one restricts 
the class of allowable solutions to those satisfying these conditions. This 
is the case of radiation fields (plane waves). Yet a linearized exact 
solution is merely of the form $g_{\mu\nu}= \eta_{\mu\nu} + h_{\mu\nu}$ 
with $|h_{\mu\nu}|\ll 1$ and no restrictions on the derivatives. For example, 
if $h_{\mu\nu}= O(\frac{1}{r})$ then $|h_{\mu\nu,\alpha}|= O(h^2)$ etc. 
and a weak--field approximation may not be a solution of (A.1). This occurs 
for the linearized Schwarzschild solution in the standard coordinates, 
\begin{displaymath}
\ud s^2= -(1-\frac{2M}{r})\,\ud t^2 
+(1+\frac{2M}{r})\,\ud r^2 +r^2\ud \Omega^2,
\end{displaymath} 
for $M/r\ll 1$; the linearized Einstein tensor ( with no gauge imposed) does 
not vanish, $G^L_{00}(h)= 12M/r^3$. On the other hand, once a differential 
equation has been generated it forgets the conditions under which it was 
derived and the space of solutions is determined solely by its form. Thus 
eq. (A.1) in vacuum has no static solutions which are globally bounded and 
the conditions for the derivatives cannot hold. Yet for the linearized 
Schwarzschild metric expressed in the isotropic coordinates, 
\begin{displaymath}
\ud s^2= -(1-\frac{2M}{\bar{r}})\,\ud t^2 
+(1+\frac{2M}{\bar{r}})(\ud \bar{r}^2 +\bar{r}^2\ud \Omega^2),
\end{displaymath} 
the perturbations satisfy the harmonic gauge condition and are a solution 
to eq. (A.1), clearly they give rise to the Newtonian acceleration 
$1/\bar{r}^{2}$. 

\section*{Appendix B. The issue of a Newtonian limit in de Sitter and AdS spaces}
Here we discuss physical and geometrical arguments showing that no Newtonian limit 
does exist in the rigorous sense in de Sitter or anti--de Sitter space: the Newtonian 
field may be defined only 'locally', i.e.~on length scales much smaller than 
$|\Lambda|^{-1/2}$, it cannot fill the entire spacetime. 
As it was discussed in appendix A the heart of the 
problem lies in geometry the ground state spacetime, whether it admits 
a foliation and a coordinate system giving rise to a structure which is 
close to Galilei spacetime.\\ 

De Sitter space is globally dynamically stable in general relativity. 
Yet its geometrical structure does not allow to define the Newtonian 
limit in the framework of Ehlers' frame theory. In fact, in the 
literature 
there are known eleven families of coordinate systems exhibiting 
various features of dS geometry and these can be divided into three 
groups corresponding to three distinct foliations of the spacetime 
\cite{SBK}.\\
i) Standard cosmological coordinates. The spacetime is sliced with 
spacelike 3--spheres $S^3$ which are O(4) invariant. The coordinate 
system is global (covers the entire manifold) and the spaces are 
almost exponentially expanding (or contracting) in the proper time 
of the observers at rest. The metric exhibits an everywhere timelike 
conformal Killing vector. A foliation by 3--spheres can also be done 
in terms of static coordinates (the metric is time--independent), 
which cover only a half of dS (the spaces consist of two hemispheres). 
The static coordinates make explicit the hypersurface orthogonal 
Killing vector which is timelike only within the cosmological event 
horizon.\\
ii) The flat cosmological coordinates which are global if the 
conformal time is employed. The spaces are just flat euclidean spaces 
which are E(3) invariant. The spatial metric is conformal--time 
dependent and this time coordinate defines a conformal timelike 
Killing field. Another slicing by flat hyperplanes may be introduced 
in a region of dS manifold using conformally Minkowski coordinates. 
The metric is also time--dependent and no (conformal) Killing vector 
is generated by this time coordinate. \\
iii) The open (hyperbolic) coordinates. The foliating spacelike 
hypersurfaces are isometric to the homogeneous Lobatchevski space 
$H^3$ of constant negative curvature which is O(1, 3) invariant. The 
coordinates cover only a half of dS. The hyperboloids $H^3$ expand or 
contract almost exponentially in the time variable which generates a 
conformal timelike Killing field.\\
None of these foliations is in general superior to the others and none 
of them is compatible with Galilei spacetime and thus it is clear that 
the frame theory cannot provide a satisfactory notion of Newtonian limit 
for de Sitter space.\\

The case of anti--de Sitter space is distinct. This manifold has 
topology $\mathbf{R}^4$ and is globally static and though is not 
globally hyperbolic, at first sight it should be more likely to have a 
structure close to Galilei spacetime. Also this manifold does not admit 
a foliation by static spacelike hyperplanes, but the main difficulty 
lies in properties of motion of test particles and metric perturbations. 
Contrary to the case of dS spacetime, AdS does admit a natural global 
static decomposition into space and time,
\renewcommand{\theequation}{\Alph{section}.\arabic{equation}} 
\setcounter{section}{2}
\setcounter{equation}{0} 
\begin{equation}
\ud s^2= 
a^2[-\cosh^2 r\,\ud t^2 +\ud r^2 +\sinh^2 r(\ud \theta^2+
\sin^2\theta\, \ud \varphi^2)],
\end{equation}
(the cosmological constant is $\Lambda=-3/a^2$) where the 
Lobatchevski hyperboloids $H^3$ given by $t=\textrm{const}$ are orthogonal 
to the globally timelike Killing vector $\partial/\partial t$ and the 
radial coordinate $r$ is distinguished among many radial variables in 
$H^3$ by the feature that it directly measures the distance along 
spatial radial geodesic lines, $s=ar$. However the hyperboloids, being 
spaces of constant curvature, do not flatten under the linearization 
($r\to 0$ or $r\to \infty$). Further, test particles behave in a rather 
bizarre way in this spacetime. A particle with an initial position 
$r=r_0>0$ (there is a coordinate singularity at the centre $r=0$ while 
$H^3$ is a homogeneous space) cannot escape to the spatial infinity 
$r=\infty$ and it cannot remain at rest even if its initial 
three--velocity is zero. The particle follows a radial geodesic and 
falls towards the centre and then recedes farther in the opposite 
direction until reaches the point $r=r_0$ at the distance 
 \begin{displaymath}
L= 2a\ln\left(Ea+\sqrt{(Ea)^2-1}\right),
\end{displaymath} 
where $E$ is the integral of energy for the geodesic line subject to 
$\cosh r_0=Ea>1$. Then the particle falls down back and returns to the 
starting point. In other terms the particle oscillates between the 
opposite points at $r=r_0$ like a pendulum. The period of these 
oscillations is universal\footnote{This periodicity in the covering 
AdS is a residual effect of the time periodicity of the original AdS 
space containing closed timelike curves.} (i.e.~is independent of 
$r_0$) and is $2\pi a$ in the proper time $s$ and $2\pi$ in the 
coordinate time $t$. The three--velocity of the particle has modulus 
equal to (in units $c=1$)
 \begin{displaymath}
[1-(Ea)^{-2}\cosh^2 r]^{1/2}
\end{displaymath} 
and not far from the centre the velocity becomes relativistic. The 
fundamental reason for these bizarre features of test particle motion 
in the covering AdS space is that no analogue of the Hopf--Rinow 
theorem for Riemannian manifolds exists for Lorentzian spacetimes and 
that AdS space is not globally hyperbolic \cite{BEE}. Hence for two 
test particles, each performing this kind of motion, it is very hard 
to define the Newtonian interaction.\\
This is, however, not the end of the story. AdS space is globally 
linearization stable \cite{IW} and nonlinearly stable for finite time 
\cite{F2} (at present it is only believed that it is globally nonlinearly 
stable). Yet if a spacetime is weakly asymptotic to the exact AdS space 
to the infinite past and future, then it is globally isometric to the 
exact AdS spacetime \cite{An2}. A regular (i.e. no singularities) 
perturbation of AdS remains close to it for long time (or possibly 
globally) but cannot tend to this spacetime at the infinity. Perturbations 
in AdS neither disappear at the infinity nor tend to a stationary 
perturbation in a far future, they are for ever travelling through the 
background. This is in marked contrast to the $\Lambda =0$ case where 
small global perturbations of Minkowski space disperse in time and 
asymptotically tend to this spacetime. In flat spacetime the Newtonian 
interaction of a system of massive bodies can be unambiguously defined 
because if in a distant region of space a gravitational perturbation 
arises, it passes through the system in a finite time interval and then 
fades away at the infinity. Yet in AdS space the perturbation will be 
present for ever and inextricably disturb interactions between the 
bodies. \\

The corollary is that dS space due to its geometrical structure and AdS 
space due to both its geometrical structure, test particle motion and the 
behaviour of the gravitational perturbations (in particular the 
non--existence of stationary excitations vanishing at the infinity), do 
not admit the Newtonian limit in the strict sense of the notion in the 
framework of general relativity with $\Lambda \neq 0$.

\end{document}